\documentclass{article}%
\usepackage{amsfonts}
\usepackage{amsmath}
\usepackage{amssymb}
\usepackage{graphicx}%
\begin{document}

\title{On the differential geometry of curves in Minkowski space}
\author{J. B. Formiga and C. Romero\\Departamento de F\'{\i}sica, Universidade Federal da Para\'{\i}ba,\\C.Postal 5008, 58051-970 Jo\~{a}o Pessoa, Pb, Brazil\\E-mail: cromero@fisica.ufpb.br }
\maketitle

\begin{abstract}
We discuss some aspects of the differential geometry of curves in Minkowski
space. We \ establish the Serret-Frenet equations in Minkowski space and use
them to give a very simple proof of the fundamental theorem of curves in
Minkowski space. We also state and prove two other theorems which represent
Minkowskian versions of a very known theorem of the differential geometry of
curves in tridimensional Euclidean space. We discuss the general solution for
torsionless paths in Minkowki space. We then\ apply the four-dimensional
Serret-Frenet equations to describe the motion of a charged test particle in a
constant and uniform electromagnetic field and show how the curvature and the
torsions of the four-dimensional path of the particle contain information on
the electromagnetic field acting on the particle.

\end{abstract}

\section{Introduction}

Our aim in this paper is to look at some aspects of the differential geometry
of curves in Minkowski space \cite{Naber}. The article is organized as
follows. In Section 2 we start by setting up the Serret-Frenet equations
in\ Minkowski space. These equations, originally formulated in Euclidean space
$R^{3},$ constitute a beautiful set of vector differential equations which
contains all intrinsic properties of a parametrized curve \cite{Manfredo}. As
is well known, this set of equations gives the derivatives with respect to the
arc length parameter of the tangent, normal and binormal vectors of a curve in
terms of each other. \ Through the Serret-Frenet equations, the evolution of a
curve is completely determined, up to rigid motion, by two intrinsic scalars:
the curvature $k(s)$ and the torsion $\tau(s).$ This result is known as the
fundamental theorem of space curves \cite{Manfredo}. As it would be expected,
an analogous theorem holds in Minkowski space, a simple proof of which is
given in Section 3. Another known theorem from the local theory of curves in
$R^{3}$ states that a curve lies in a plane if and only if its torsion
vanishes \cite{Manfredo}, a result that gives us clear geometrical insight in
the notion of torsion. This theorem has two counterparts in Minkowski space
and are also discussed in section 3. In section 4 we illustrate the formalism
developed previously with a discussion of the motion of a relativistic charged
particle placed in external electromagnetic field. As a direct application of
the ideas developed in section 2 and section 3 we obtain, in Section 5, the
general solution of the Serret-Frenet equations for torsionless paths,\ which
may be viewed as corresponding to a class of accelerated observers, including,
as a particular case, the so-called Rindler observers. Finally, in Section 6,
we give a simple derivation of the connection between the curvature and
torsions of the worldline of a charged particle placed in an external
electromagnetic field and the relativistic invariants of the field.

\bigskip

\section{The Serret-Frenet equations}

In Euclidian space $R^{3}$ the intrinsic geometric properties of a curve
$\Gamma$ (parametrized by the arc length $s$) are described by the
Serret-Frenet equations
\begin{align}
\frac{dt}{ds}  &  =kn\label{Serret}\\
\frac{dn}{ds}  &  =-kt+\tau b\nonumber\\
\frac{db}{ds}  &  =-\tau n\nonumber
\end{align}
or, in matrix representation,
\begin{equation}%
\begin{bmatrix}
\frac{dt}{ds}\\
\frac{dn}{ds}\\
\frac{db}{ds}%
\end{bmatrix}
=%
\begin{bmatrix}
0 & k & 0\\
-k & 0 & \tau\\
0 & -\tau & 0
\end{bmatrix}%
\begin{bmatrix}
t\\
n\\
b
\end{bmatrix}
\label{Serret2}%
\end{equation}
where $t,n,b$ denote, respectively, the tangent, normal and binormal vectors
of the curve $\Gamma$, which is assumed to be smooth \ (at least of class
$C^{3}$ ). The triad of vectors $(t,n,b)$ constitute an orthonormal
right-handed frame defined at each point of $\Gamma$ and the invariant scalars
$k=k(s)$ and $\tau_{1}=\tau_{1}(s)$ are called, respectively, the curvature
and torsion of $\Gamma.$ The equations (\ref{Serret}) follow directly from the
definition of the normal vector $n$ and the binormal vector $b$ $(b\equiv
t\wedge n)$ \cite{Manfredo}

To adapt the above formalism to Minkowski space we need to replace the
Euclidian metric for the Minkowski metric $\eta_{\alpha\beta}%
=diag(1,-1,-1,-1)$\footnote{We follow the convention in which Greek indices
range from $0$ to $3$, and use the so-called Einstein summation rule, i.e. in
the case of repeated index summation over the appropriate range is
implied.}$,$ and define a \textit{second binormal} and a \textit{second
torsion }$\tau_{2}=\tau_{2}(s)$ . Since usual vector products make no sense in
four-dimensional space we define our set of orthonormal four-vectors (a
\textit{tetrad}) by concomitantly requiring them to satisfy a four-dimensional
extension of the Frenet-Serret equations, which then governs the evolution of
the tetrad. \ It is also convenient to restrict ourselves to timelike curves
$x^{\alpha}=x^{\alpha}(s)$, i.e. those for which $\eta_{\alpha\beta}%
\frac{dx^{\alpha}}{ds}\frac{dx^{\beta}}{ds}=1$, where now $s$ denotes the arc
length parameter in the sense of Minkowski metric $\eta_{\alpha\beta}.$
Accordingly, if we denote the tetrad vectors by $u_{(A)}^{\alpha}$
$(A=0,...3)$ \footnote{From now on Latin capital indices $(A,B,...)$ are
labels for distinguishing the particular vector of the tetrad, while
$(\alpha,\beta,...)$ denotes its components. For convenience, however, the
vectors of the canonical coordinate basis, will be denoted by $e_{(\alpha)}.$%
}, then the orthonormality conditions read $u_{(A)}^{\alpha}u_{(B)\alpha
}\equiv\eta_{\alpha\beta}u_{(A)}^{\alpha}u_{(B)}^{\beta}=\eta_{AB}.$ It can be
shown \cite{Syngebook,Lanczos} that if we chose $u_{(0)}^{\alpha}%
=\frac{dx^{\alpha}}{ds}$, i;e. $u_{(0)}^{\alpha}$ being the components of the
unit tangent vector , then we can easily construct an orthonormal basis of
vectors $\{u_{(A)}^{\alpha}\}$ ,defined along the curve, which obey the
following four-dimensional Serret-Frenet equations, given in matrix
representation by \footnote{We know that in Minkowski space vectors which are
orthogonal to timelike vectors are spacelike. Thus, as $\ \frac{du_{(0)}}{ds}$
is orthogonal to $u_{(0)},$ then $\frac{du_{(0)}}{ds}$ is spacelike. We
conveniently define $k(s)$ to be positive and given by $\ k(s)$ $\equiv(-$
$\eta_{\alpha\beta}\frac{du_{(0)}^{\alpha}}{ds}\frac{du_{(0)}^{\beta}}%
{ds})^{-1/2}.$ We also conventionally define $\tau_{1}$ to be non-negative.}
\begin{equation}%
\begin{bmatrix}
\frac{du_{(0)}}{ds}\\
\frac{du_{(1)}}{ds}\\
\frac{du_{(2)}}{ds}\\
\frac{du_{(3)}}{ds}%
\end{bmatrix}
=%
\begin{bmatrix}
0 & k & 0 & 0\\
k & 0 & \tau_{1} & 0\\
0 & -\tau_{1} & 0 & \tau_{2}\\
0 & 0 & -\tau_{2} & 0
\end{bmatrix}%
\begin{bmatrix}
u_{(0)}\\
u_{(1)}\\
u_{(2)}\\
u_{(3)}%
\end{bmatrix}
\label{Serret4}%
\end{equation}
Of course the above procedure may be easily generalized to $n$-dimensional
Riemannian (or pseudo-Riemanian) spaces by changing from ordinary
differentiation to absolute differentiation \cite{Syngebook,Lanczos}. Here,
two points are worth mentioning. First, due to the Lorentzian signature, the
$4\times4$ matrix that governs the evolution of the tetrad vectors
$u_{(A)}^{\alpha}$ is not anti-symmetric, as in the case of Euclidean
signature. Secondly, in order to construct of the tetrad $\{u_{(A)}^{\alpha
}\}$ it is not necessary, as a matter of fact, to assume that the curvature
and the torsions have non-zero values. If $k=0$, then the curve is a timelike
geodesic and a triad of constant spacelike orthonormal vectors $\{u_{(1)}%
,u_{(2)},u_{(3)}\}$ orthogonal to $u_{(0)}$ may be chosen. In this case,
$\tau_{1}(s)$ and $\tau_{2}(s)$ are zero. If $k\neq0$, but $\tau_{1}(s)$ $=0$,
then we can choose an orthonormal basis\ $\{u_{(0),}$ $u_{(1)},u_{(2)}%
,u_{(3)}\}$ in such a way that $u_{(2)}$ and $u_{(3)}$ are constant spacelike
vectors (see (\cite{Syngebook}) for details).

\section{\bigskip The fundamental theorem in Minkowski space}

A most important result in the local theory of curves in Euclidean space
$R^{3}$, known as the fundamental theorem of curves, states the following:

\textbf{Theorem 1} \textit{Given differentiable functions }$k(s)$ $>0$\textit{
and }$\tau(s)$\textit{, there exists a regular parametrized curve} $\Gamma$
\textit{such that }$k(s)$\textit{ is the curvature, and }$\tau(s)$\textit{ is
the torsion of \ }$\Gamma.$\textit{ Any other curve }$\overline{\Gamma}%
$\textit{ satisfying the same conditions, differs from }$\Gamma$\textit{ by a
rigid motion} \cite{Manfredo}\textit{. }It would be natural to expect this
theorem to hold when appropriately transposed to Minkowski space. In this
case, a \textit{rigid motion} would correspond to a Poincar\'{e}
transformation\textit{ \footnote{It is usual to consider a Poincar\'{e}
transformation as coordinate transformation in the passive sense. Here, we
regard it in the active sense, i.e. as a \textit{motion }in Minkowski space.}
}and the curve would be expected to be determined by the three differentiable
functions $k(s)$ $>0$, $\tau_{1}(s)$ and $\tau_{2}(s).$ In this section we
give a simple proof of the \textit{fundamental theorem of curves in Minkowski
space}. We omit the proof of the existence part since it is almost the same as
in the case of $R^{3}$, requiring only minor modifications (see, for instance,
the appendix to Chap. 4 of \cite{Manfredo} )\textit{.} The proof of the
uniqueness part, however, differs from its counterpart in $R^{3\text{ }}$,
since the latter makes use of the positiveness of the Euclidean metric. Let us
first state the theorem.

\textbf{Theorem 1* \ }\textit{Given} \textit{differentiable functions }$k(s)$
$>0$, $\tau_{1}(s)$ \textit{and} $\tau_{2}(s)$\textit{, there exists a regular
parametrized timelike \footnote{In developing the Serret-Frenet formalism in
Minkowski space we are ultimately interested in applications to the motion of
physical particles. Therefore we shall restrict ourselves to timelike curves,
although the formalism may also work for spacelike curves. Consideration of
null curves should introduce some difficulty.}\ curve }$\Gamma$ \textit{such
that }$k(s)$ \textit{is the curvature, }$\tau_{1}(s)$ \textit{and }$\tau
_{2}(s)$ \textit{are, respectively, the first and second torsion of \ }%
$\Gamma$. \textit{Any other curve }$\overline{\Gamma}$\textit{ satisfying the
same conditions, differs from }$\Gamma$ \textit{by a Poincar\'{e}
transformation}, \textit{i.e. by a transformation of the type }$x^{\prime\mu
}=\Lambda_{..\nu}^{\mu}x^{\nu}$\textit{ }$+$\textit{ }$a^{\mu}$\textit{ ,
where }$\Lambda_{..\nu}^{\mu}$\textit{ represents a proper Lorentz matrix and
}$a^{\mu}$\textit{ is a constant four-vector \footnote{A matrix $\Lambda$ is
said to be a \textit{Lorentz matrix} if it is pseudo-ortogonal, i.e. \ if it
satisfies the condition $\Lambda^{t}\eta\Lambda=\eta$, where $\Lambda^{t}$ is
the transposed of $\Lambda$, and $\eta$ denotes the Minkowski metric. If
$\det\Lambda=1$, then $\Lambda$ is called a \textit{proper Lorentz matrix.}}.
\ }

We now give a proof of the uniqueness, up to a Poincar\'{e} transformation, of
the above result. Let us assume that two timelike curves $\Gamma$ and
$\overline{\Gamma}$ satisfy the conditions $k(s)$ $=\overline{k}(s)$,
$\tau_{1}(s)=\overline{\tau}_{1}(s)$ and $\tau_{2}(s)=\overline{\tau}_{2}(s)$,
with $s\in I$, where $I$ is an open interval of $R.$ Let $\{u_{(A)}^{\alpha
}(s_{0})\}$, $\{\overline{u}_{(A)}^{a}(s_{0})\}$ be the Serret-Frenet tetrads
at $s_{0}\in I$ of $\Gamma$ and $\overline{\Gamma}$, respectively. It is clear
that it is always possible, by a Poincar\'{e} transformation, to bring
$x^{\mu}(s_{0})$ of $\Gamma$ into $\overline{x}^{\mu}(s_{0})$ of
$\overline{\Gamma}$ in such a way that $u_{(A)}^{\alpha}(s_{0})=\overline
{u}_{(A)}^{a}(s_{0})$. Now, the two Serret-Frenet tetrads $\{u_{(A)}^{\alpha
}(s)\}$, $\{\overline{u}_{(A)}^{\alpha}(s)\}$ satisfy the equations
\begin{equation}%
\begin{bmatrix}
\frac{du_{(0)}}{ds}\\
\frac{du_{(1)}}{ds}\\
\frac{du_{(2)}}{ds}\\
\frac{du_{(3)}}{ds}%
\end{bmatrix}
=%
\begin{bmatrix}
0 & k & 0 & 0\\
k & 0 & \tau_{1} & 0\\
0 & -\tau_{1} & 0 & \tau_{2}\\
0 & 0 & -\tau_{2} & 0
\end{bmatrix}%
\begin{bmatrix}
u_{(0)}\\
u_{(1)}\\
u_{(2)}\\
u_{(3)}%
\end{bmatrix}
\end{equation}
and
\begin{equation}%
\begin{bmatrix}
\frac{d\overline{u}_{(0)}}{ds}\\
\frac{d\overline{u}_{(1)}}{ds}\\
\frac{d\overline{u}_{(2)}}{ds}\\
\frac{d\overline{u}_{(3)}}{ds}%
\end{bmatrix}
=%
\begin{bmatrix}
0 & k & 0 & 0\\
k & 0 & \tau_{1} & 0\\
0 & -\tau_{1} & 0 & \tau_{2}\\
0 & 0 & -\tau_{2} & 0
\end{bmatrix}%
\begin{bmatrix}
\overline{u}_{0}\\
\overline{u}_{1}\\
\overline{u}_{2}\\
\overline{u}_{3}%
\end{bmatrix}
\end{equation}
which can be written in a more compact form as
\begin{align}
\frac{du_{(A)}}{ds}  &  =\Sigma_{A}^{..B}u_{(B)}\label{sigma}\\
\frac{d\overline{u}_{(A)}}{ds}  &  =\Sigma_{A}^{..B}\overline{u}%
_{(B)}\nonumber
\end{align}
with $\Sigma_{A}^{..B}$denoting the\ elements of the Serret-Frenet matrix.
Clearly, the two tetrads $\{u_{(A)}(s)\}$, $\{\overline{u}_{(A)}(s)\}$ are
related by an equation of the type
\begin{equation}
\overline{u}_{(A)}(s)=\Lambda_{A}^{..B}(s)u_{B}(s) \label{lambda}%
\end{equation}
with the elements of the matrix $\Lambda(s)$ satisfying the condition
$\Lambda_{A}^{..B}(s_{0})=\delta_{B}^{A}$ , since we are assuming that
$u_{(A)}^{\alpha}(s_{0})=\overline{u}_{(A)}^{a}(s_{0})$. From (\ref{sigma})
and (\ref{lambda}) we obtain a system of first-order differential equations
for the elements of $\Lambda$ given by
\begin{equation}
\frac{d\Lambda_{A}^{..B}}{ds}+\Lambda_{A}^{..C}\Sigma_{C}^{..B}-\Sigma
_{A}^{..C}\Lambda_{C}^{..B}=0 \label{equation1}%
\end{equation}
By assumption, $\Sigma_{A}^{..B}$ are differentiable functions of the proper
parameter $s.$ From the theory of ordinary differential equations \cite{Lang},
we know that if we are given a set of initial conditions\textit{ }$\Lambda
_{A}^{..B}(s_{0})$, then the above system admits a unique solution
$\{\Lambda_{A}^{..B}=\Lambda_{A}^{..B}(s)\}$\ defined in a open interval
$J\subset I$ containing $s_{0}.$ On the other hand, it is easily seen that
$\Lambda_{A}^{..B}(s)=\delta_{B}^{A}$ is a solution of (\ref{equation1}).
Therefore, we conclude that $\overline{u}_{(A)}(s)=u_{A}(s).$

Other extensions of known theorems of the differential geometry of curves in
Euclidian space $R^{3}$ can easily be carried over into Minkowski space. As an
example, let us consider the following results on curves lying in $R^{3}%
$:\textbf{\ }

\textbf{Theorem 2 }\textit{A curve, with non-vanishing curvature, is plane if
and only if its torsion vanishes identically }\cite{Kreyszig}\textit{.
}Natural extensions of this result to Minkowski space are given by the
following theorems:

\textbf{Theorem 2* \ }\textit{A timelike curve }$\Gamma$\textit{, with
non-vanishing curvature, lies in a hyperplane if and only if the second
torsion vanishes identically. }

\textit{Proof.} Again we restrict ourselves to timelike curves. Let us start
with the necessary condition. Suppose the curve $\Gamma$\ lies in a
hyperplane. Then, by a Lorentz rotation we can align one of the coordinate
axes with the normal direction to the hyperplane. For the sake of the
argument, let us assume that we can bring $\Gamma$\ to lie, say, in the
$(x^{0},x^{1},x^{2})$-hyperplane. Then the parametric equations of $\Gamma$
are of the form $x^{\alpha}=x^{\alpha}(s)=(x^{0}(s),x^{1}(s),x^{2}(s),0)$. Let
$\{e_{(\alpha)}\}$ denote the vectors of the \ canonical coordinate basis.
Thus, in these coordinates, $u_{(0)}=\frac{dx^{0}}{ds}e_{(0)}+\frac{dx^{1}%
}{ds}e_{(1)}+\frac{dx^{2}}{ds}e_{(2)}$ and $\ \frac{du_{(0)}}{ds}=\frac
{d^{2}x^{0}}{ds^{2}}e_{(0)}+\frac{d^{2}x^{1}}{ds^{2}}e_{(1)}+\frac{d^{2}x^{2}%
}{ds^{2}}e_{(2)}.$ From (\ref{Serret4}) we have $\frac{du_{(0)}}{ds}=ku_{(1)}%
$. Given that $k\neq0$ we conclude that $u_{(1)}$ has no components in the
$x^{3}$-direction, i.e., $u_{(1)}=(u_{(1)}^{0},u_{(1)}^{1},u_{(1)}^{2},0).$
Thus, $\frac{du_{(1)}}{ds}=$\ $\frac{du_{(1)}^{0}}{ds}e_{(0)}+\frac
{du_{(1)}^{1}}{ds}e_{(1)}+\frac{du_{(1)}^{2}}{ds}e_{(2)}$, hence from the
equation $\frac{du_{(1)}}{ds}=ku_{(0)}+\tau_{1}u_{(2)}$ we conclude that
$\tau_{1}u_{(2)}^{3}=0.$ If $\tau_{1}=0$, then $\tau_{2}$ also must vanish,
for in this case $u_{(2)}$ is chosen to be constant. If $\tau_{1}\neq0$, then
$u_{(2)}^{3}=0,$ hence $u_{(2)}=(u_{(2)}^{0},u_{(2)}^{1},u_{(2)}^{2},0)$. From
the equation $\frac{du_{(2)}}{ds}=$\ $\frac{du_{(2)}^{0}}{ds}e_{(0)}%
+\frac{du_{(2)}^{1}}{ds}e_{(1)}+\frac{du_{(2)}^{2}}{ds}e_{(2)}$ and the third
Serret-Frenet equation $\frac{du_{(2)}}{ds}=-\tau_{1}u_{(1)}+\tau_{2}u_{(3)}$
we are led to conclude that $\tau_{2}u_{(3)}^{3}=0$. However, $u_{(3)}^{3}$
cannot be zero, otherwise the set of vectors $\{u_{(0)},u_{(1)},u_{(2)}%
,u_{(3)}\}$ would not be linearly independent. Therefore, $\tau_{(2)}$ must vanish.

Let us turn to the sufficient condition. Suppose that $\tau_{2}=0.$ Then, the
fourth Serret-Frenet equation implies that $u_{(3)}$ is a \ constant vector.
Let us conveniently choose our coordinate system in such a way that $e_{(3)}=$
$u_{(3)}$. Now, since $u_{(0)}$ is orthogonal to $u_{(3)}$ we must have
$u_{(0)}^{3}=0$, which means that $\Gamma$ lies in the hyperplane
$x^{3}=const.$ This completes the proof. Another extension of Theorem 2 leads
directly to the following proposition:

\textbf{Theorem2** \ }\textit{A timelike curve }$\Gamma$\textit{, with
non-vanishing curvature, is plane \ if and only if the first and second
torsions vanishes identically. }

We omit the proof since it follows the same lines of reasoning presented above.

\section{\bigskip Charged particles in electromagnetic fields}

The Serret-Frenet formalism adapted to four-dimensional Lorentzian spaces may
be very useful in providing geometric insight into the motion of accelerated
particles, both in the context of special relativity and general relativity
\cite{Carmeli,Synge,Pina,Honig,Ringermacher, Bonilla} . Synge \cite{Synge},
for instance, applied the formalism to investigate intrinsic geometric
properties of the world lines of charged particles placed in an
electromagnetic field, showing that for a constant and uniform electromagnetic
field the point charge describes a timelike helice in Minkowski space. Another
interesting result was obtained in \cite{Honig}, in which the authors
establish a connection between the intrinsic scalars associated with the
timelike curve of a charged particle in a constant and uniform electromagnetic
field and the field invariants. In this section, with the aid of the formalism
developed in Section 2, we explore some physical and geometrical consequences
of the connection mentioned above. Also, as a direct application of
Theorem2**, we shall find the general solution of the motion of a class of
accelerated observers, which includes, as a particular case, the well-known
Rindler observers.

Let us start by considering the motion of a charged particle of rest mass $m$
and charge $e$ placed in an external electromagnetic field in Minkowski space.
The motion of the particle is governed by the Lorentz equation%
\begin{equation}
\frac{du_{(0)}^{\alpha}}{ds}=\frac{e}{mc^{2}}F_{..\beta}^{\alpha}%
u_{(0)}^{\beta} \label{Lorentz}%
\end{equation}
where $u_{(0)}$ is the 4-velocity of the particle, $F_{\alpha\beta}$ is the
electromagnetic field tensor and $c$ is the speed of light in vacuum
\cite{Landau}. Let us now consider the Serret-Frenet formalism and look at
(\ref{Lorentz}) as one of the four equations governing the motion of the
tetrad defined by (\ref{Serret4}). When $F_{\alpha\beta}$ is uniform and
constant, it can be shown that the three Serret-Frenet scalars remain constant
along the world line of the particle, i.e. $\frac{dk}{ds}=\frac{d\tau_{1}}%
{ds}=\frac{d\tau_{2}}{ds}=0$ \cite{Honig}. Moreover,\ in this case the two
invariants of the electromagnetic field$\ \ I_{1}=F_{\alpha\beta}%
F^{\alpha\beta}=2(B^{2}-E^{2})$ and $I_{2}=F_{\alpha\beta}^{\ast}%
F^{\alpha\beta}=4\overrightarrow{E}.\overrightarrow{B}$ are related to the
Frenet-Serret scalars \cite{Honig} through the equations
\begin{equation}
B^{2}-E^{2}=\frac{m^{2}c^{4}}{e^{2}}(\tau_{1}^{2}+\tau_{2}^{2}-k^{2})
\label{invariant1}%
\end{equation}%
\begin{equation}
\overrightarrow{E}.\overrightarrow{B}=-m^{2}c^{4}\frac{k\tau_{2}}{e^{2}}
\label{invariant2}%
\end{equation}
Let us now consider simple examples of how these equations allow us to extract
useful information on the relationship between the geometry of the curve and
the underlying physics which determines the motion. Suppose, for instance,
that there exists a reference frame in which the magnetic field
$\overrightarrow{B}$ vanishes. Then, from (\ref{invariant2}), either $k$
$=\tau_{2}=0$ or $k$ $\neq0$ and $\tau_{2}=0$. However, from (\ref{invariant1}%
) we see that the first invariant must be negative, hence $k$ cannot vanish
and, thus, we must have $\tau_{2}=0$. Then, from theorem 2*, it follows that
the path must lie in a hyperplane.

Now suppose that the curve has curvature $(k\neq0)$, but it is torsionless,
i.e. $\tau_{1}$ $=\tau_{2}=0$. Then, from (\ref{invariant1}) and
(\ref{invariant2}) we have $\overrightarrow{E}.\overrightarrow{B}=0$ and
$B^{2}-E^{2}<0.$ Therefore, there exists a reference frame $S^{\prime}$ in
which the magnetic field $\overrightarrow{B^{\prime}}$ vanishes \cite{Landau}.
Then, we must have $k=\frac{e}{mc^{2}}\left\vert \overrightarrow{E}^{\prime
}\right\vert $, where $\overrightarrow{E}^{\prime}$ is the electric field
measured in $S^{\prime}$.

Finally, if $k=\tau_{1}\neq0$ and $\tau_{2}=0$, then the two invariants
$B^{2}-E^{2}$ and $\overrightarrow{E}.\overrightarrow{B}$ vanish, which means
that we are in the presence of a null electromagnetic field, that is, a pure
radiation field \cite{Stephani}.

\section{General solutions for torsionless paths}

Let us investigate with more details the case of torsionless timelike paths in
Minkwoski space. For the sake of simplicity, let us choose our coordinate
system such that the motion takes place in the $(x^{0},x^{1})$-plane. Then,
the Serret-Frenet equations yield%

\begin{align}
\frac{du_{(0)}}{ds}  &  =ku_{(1)}\label{planemotion}\\
\frac{du_{(1)}}{ds}  &  =ku_{(0)}\nonumber\\
\frac{du_{(2)}}{ds}  &  =\frac{du_{(3)}}{ds}=0\nonumber
\end{align}
Choosing $u_{(2)}$ and $u_{(3)}$ as the usual unit vectors $e_{(2)}$ and
$e_{(3)}$ in the $x^{2},x^{3}$- directions, respectively, it remains to solve
the two-dimensional system of differential equations for $u_{(0)}$ and
$u_{(1)}$. As it can easily be verified, the general solution of
(\ref{planemotion}) is given by
\begin{equation}
x^{\alpha}(s)=\left(
{\displaystyle\int_{0}^{s}}
\cosh(\theta(s))ds+a,\int_{0}^{s}\sinh(\theta(s))ds+b,0,0\right)
\label{general}%
\end{equation}
where the function $\theta(s)$ is given in terms of the curvature $k=$ $k(s)$
by $\theta(s)=\int_{0}^{s}k(s)ds$ + $\varphi$, with $a,$ $b$ and $\varphi$
being arbitrary constants.

Let us consider the case of constant curvature. If we choose, as initial
conditions of the motion, $x^{\alpha}(0)=(0,1,0,0)$ and $u_{(0)}^{\alpha
}(s)=(1,0,0,0)$ then from (\ref{general}) we have $x^{\alpha}(s)=\left(
\cosh(ks),\sinh(ks),0,0\right)  $, which describes a hyperbolic motion in
Minkowski space, and may be interpreted as a Rindler observer, i.e. an
accelerated particle whose proper acceleration (i.e. acceleration relative to
its instantaneous rest frame) is constant \cite{Rindler}.

\bigskip

\section{Final remarks}

Due to the fundamental theorem of curves (which is extendable to Minkowski
space as well as to more general Riemannian \ spaces), if a curve represents
the motion of a particle, one can look at the Serret-Frenet equations as
containing complete information on the dynamics of the particle. Such
correspondence between the geometry of curves and the dynamics of particles
can be nicely explored in the context of special relativity to\ study the
intrinsic geometry of world lines in Minkowski space. In fact, compared to the
Newtonian formalism, special relativity is a\ more natural setting for a
description of motion through the Serret-Frenet equations, since the
worldlines of particles are usually parametrized by the arc length parameter
$s$, turning the equations into a much simpler form. As we have seen in
Section 4, an example which particularly illuminates the connection between
physical and geometrical aspects of the dynamics of accelerated particles is
given by the motion of a charge in an external eletromagnetic field. Of
special interest is the existing correlation between the intrinsec invariants
$k,\tau_{1},\tau_{2}$ of the curve and the field invariants $I_{1}%
=F_{\alpha\beta}F^{\alpha\beta}$ and $I_{2}=F_{\alpha\beta}^{\ast}%
F^{\alpha\beta}$. Now, if we calculate the characteristic polynomial of the
Serret-Frenet matrix $\Sigma_{A}^{..B}$\ (\ref{Serret4}), we get
\begin{equation}
\lambda^{4}-(k^{2}-\tau_{1}^{2}-\tau_{2}^{2})\lambda^{2}-k^{2}\tau_{2}^{2}=0
\label{polynomial}%
\end{equation}
Thus, the special combinations of the curvature and torsions that appears in
the equations (\ref{invariant1}) and (\ref{invariant2}) are no more than two
of the coefficients of the characteristic polynomial associated with the
Serret-Frenet matrix. On the other hand, it is well known that the
characteristic polynomial corresponding to the electromagnetic tensor
$F_{\alpha\beta}$ is given by \cite{Gerardo}%

\begin{equation}
\lambda^{4}+\frac{I_{1}}{2}\lambda^{2}-\left(  \frac{I_{2}}{4}\right)  ^{2}=0
\label{polynomial3}%
\end{equation}

From this findings we can derive the equations (\ref{invariant1}) and
(\ref{invariant2}), obtained in ref. \cite{Honig}, in a different and easier
way. First, let us note that, if the eletromagnetic field tensor is constant
and uniform, it can be shown, by combining the equations (\ref{Serret4}) and
(\ref{Lorentz}), that the following equation holds \cite{Honig} :%
\begin{equation}
\frac{du_{(A)}^{\alpha}}{ds}=\frac{e}{mc^{2}}F_{..\beta}^{\alpha}%
u_{(A)}^{\beta} \label{Lorentz3}%
\end{equation}
Let us express the components of the tensor $F_{..\beta}^{\alpha}$\ in the
tetrad basis $\{u_{(A)}\}$. The canonical vector basis $e_{(\alpha)}$ may be
written with respect to the Serret-Frenet basis as\ $e_{(\alpha)}=$
$u_{\alpha}^{..(A)}u_{(A)}$ $.$ Then, as$\ \frac{du_{(A)}}{ds}=\frac
{du_{(A)}^{\alpha}}{ds}e_{(\alpha)}$, the equation (\ref{Lorentz3}) takes the
form
\begin{equation}
\frac{du_{(A)}}{ds}=\digamma_{..A}^{B}u_{(B)} \label{Lorentz2}%
\end{equation}
where we are defining $\digamma_{..A}^{B}\equiv\frac{e}{mc^{2}}F_{..\beta
}^{\alpha}u_{(A)}^{\beta}$ $u_{\alpha}^{..(B)}$. \ Since $\digamma
_{AB}\digamma^{AB}=\left(  \frac{e}{mc^{2}}\right)  ^{2}F_{\alpha\beta
}F^{\alpha\beta}=\left(  \frac{e}{mc^{2}}\right)  ^{2}I_{1}$ and
$\digamma_{AB}^{\ast}\digamma^{AB}=$ $\left(  \frac{e}{mc^{2}}\right)
^{2}F_{\alpha\beta}^{\ast}F^{\alpha\beta}=\left(  \frac{e}{mc^{2}}\right)
^{2}I_{2}$ ,$\ $ the characteristic polynomial associated with $F_{..B}^{A}$
is given by \footnote{Here we are raising and lowering indices of
$F_{...B}^{A}$ with the Minkowski metric $\eta_{AB}$.}
\begin{equation}
\lambda^{4}+\left(  \frac{e}{mc^{2}}\right)  ^{2}\frac{I_{1}}{2}\lambda
^{2}-\left(  \frac{e}{mc^{2}}\right)  ^{4}\left(  \frac{I_{2}}{4}\right)
^{2}=0 \label{polynomial2}%
\end{equation}

Thus, from (\ref{polynomial}) and (\ref{polynomial2}) we obtain
(\ref{invariant1}) and (\ref{invariant2}) \footnote{Actually we can infer that
$\overrightarrow{E}.\overrightarrow{B}=\pm m^{2}c^{4}\frac{k\tau_{2}}{e^{2}}$.
That the equation should hold with the minus sign is a consequence of our
having defined $k$ and $\tau_{1}$as non-negative scalars. (See ref. [10]).}.

The beautiful result proved above was obtained under the\ stringent condition
that requires $F_{\alpha\beta}$\ to be constant and uniform (or, homogeneous,
in the Riemannian case \cite{Honig}). If we drop the assumption that the
electromagnetic field is constant and uniform, then (\ref{Lorentz3}) and
(\ref{Lorentz2}) no longer hold. However, it is still possible to find a set
of equations relating the intrinsic parameters of the curve to the components
of electromagnetic field tensor and their time derivatives along the curve
written in the Serret-Frenet basis. After some algebra, it can be shown that
in the general case we have the following relations:%

\begin{equation}
k=\digamma_{01};\text{ }\tau_{1}=\digamma_{12}+\frac{G_{02}}{k};\text{ }%
\tau_{2}=\frac{1}{\tau_{1}k}\left(  \frac{\dot{k}}{k}G_{30}-H_{03}\right)
-\frac{2}{\tau_{1}}G_{13}+\digamma_{23} \label{generalcase}%
\end{equation}
where $\dot{k}=\frac{dk}{ds}$, $\digamma_{AB}=\frac{e}{mc^{2}}F_{\alpha\beta
}u_{(A)}^{\alpha}u_{(B)}^{\beta}$, $G_{AB}=\frac{e}{mc^{2}}(\frac{d}%
{ds}F_{\alpha\beta})u_{(A)}^{\alpha}u_{(B)}^{\beta}$ and $H_{03}=\frac
{e}{mc^{2}}(\frac{d^{2}}{ds^{2}}F_{\alpha\beta})u_{(0)}^{\alpha}u_{(3)}%
^{\beta}$. In the case of a constant and uniform electromagnetic field the
above equations simplifies to $k=F_{01}$, $\tau_{1}=F_{21}$and $\tau
_{2}=F_{23}$.

\bigskip

\section{Acknowledgement}

The authors would like to thank CNPq-FAPESQ (PRONEX) for financial support.

\end{document}